\documentclass[aps,pre,preprint,groupedaddress]{revtex4-1}

\usepackage{array}
\usepackage{booktabs}
\usepackage{multirow}
\usepackage{rotating}
\usepackage{amssymb}

\begin{document}


\title{Remodelling of the fibre-aggregate structure of collagen gels by cancer-associated fibroblasts: a time-resolved grey-tone image analysis based on stochastic modelling} 

\author{Cedric J. Gommes$^{1}$} \email{cedric.gommes@uliege.be}
\author{Thomas Louis$^{2}$}
\author{Isabelle Bourgot$^{2}$}
\author{Erik Maquoi$^{2}$}
\author{Silvia Blacher$^{2}$}
\author{Agn\`es No\"el$^{2}$}

\affiliation{$^{1}$Department of Chemical Engineering, University of Li\`ege B6A, All\'ee du Six Ao\^ut 3, B-4000 Li\`ege, Belgium}
\affiliation{$^{2}$Laboratory of Tumor and Development Biology, GIGA-Cancer, University of Li\`ege, Li\`ege, Belgium }

\begin{abstract}
Solid tumors consist of tumor cells associated with stromal and immune cells, secreted factors and extracellular matrix (ECM), that together constitute the tumor microenvironment. Among stromal cells, cancer-associated fibroblasts (CAFs) are of particular interest. CAFs modulate the architecture of the ECM by exerting forces and contracting collagen fibres, creating paths that facilitate cancer cell migration. The characterization of the collagen fibre network and its space and time-dependent microstructural modifications is key to investigating the interactions between cells and the ECM. Developing image analysis tools for that purpose is still a challenge. The structural complexity of the collagen network calls for specific statistical descriptors. Moreover, the low signal-to-noise ratio of imaging techniques available for time-resolved studies rules out standard image analysis methods. In this work, we propose an innovative data analysis approach to investigate the remodelling of fibrillar collagen in a three-dimensional spheroid model of cellular invasion, which is based on the stochastic modelling of the gel structure and on grey-tone image analysis. The method was used to study the reshaping of a collagen matrix by migrating breast cancer-derived CAFs. The structure of the gels at the scale of a few microns consists in regions with high fibre density separated by depleted regions, which can be thought of as fibre aggregates and pores. To characterize this structure, we developed a two-scale stochastic model with a clipped Gaussian field model to describe the aggregates-and-pores large-scale structure, and a homogeneous Boolean model to describe the small-scale fibre network within the aggregates. The model parameters are identified by fitting the grey-tone histograms and correlation functions of confocal microscopy images. The specificity of the method is that it applies to the unprocessed grey-tone images, and it can therefore be used with low magnification, noisy time-lapse reflectance images. When applied to the spheroid time-resolved images, the method revealed two different matrix densification mechanisms for the gel in direct contact or far from the cells.
\end{abstract}

\maketitle

 
\section{Introduction}

In living tissues, cells are commonly embedded within complex networks of extracellular matrix (ECM) constituted of diverse highly cross-linked components, including fibrous proteins, proteoglycans, glycoproteins and polysaccharides \cite{Hynes:2009,Hynes:2012,Oskarsson:2013}. Each of the individual ECM components exhibits specific biomechanical and biochemical properties that are related to its polymer structure, size and binding affinities to signaling molecules \cite{Frantz:2010,Theocharis:2016}. These properties in turn modulate several cellular processes including proliferation, differentiation, survival and motility by ligating specialized cell surface receptors \cite{Leitinger:2007,Xian:2010}. Cell-matrix interactions also lead to ECM remodeling by the cells, with the ECM fibres being synthesized, re-oriented, deformed and degraded by the cells, particularly fibroblasts \cite{Shieh:2011}. The topology of the ECM (organization of fibres, matrix porosity and density) also influences the mechanical properties of the matrix \cite{Doyle:2009,Charras:2014}. For instance, aligned bundles of collagen increase matrix stiffness \cite{Frantz:2010}. Along with fibre organization, the degree of porosity of the ECM influences its rheological properties which in turn modulates the cell behavior \cite{Seo:2020}.

Collagens are the most abundant components of the ECM and their structure and composition differ across various tissue types \cite{Yue:2014,Mouw:2014}. The collagen family is composed of 28 types consisting in fibril-forming, fibril-associated, network-forming and other structures subfamilies \cite{Bourgot:2020,Revell:2021}. Type I collagen represents the most common fibrillar collagen in vertebrates. Its synthesis, stiffening and remodelling are involved in both physiological processes such as wound healing, inflammation, tissue repair and pathological ones such as angiogenesis or tumor fibrosis which promotes growth and invasion \cite{Provenzano:2009,Provenzano:2008,Buchmann:2021,Bourgot:2020}. Many solid tumors such as those of the breast, lung and pancreas are characterized by the presence of a desmoplastic stroma typified by an accumulation of fibrillar collagen. This collagen accretion increases the stiffness and creates discrete structural patterns called tumor-associated collagen signatures (TACS) in the stroma. These anisotropic patterns promote directed cell migration into and through the stroma by contact guidance \cite{Dickinson:1994,Provenzano:2006,Ray:2021,Ray:2022}. The stroma comprises a complex ecosystem of endothelial cells, immune cells, and cancer-associated fibroblasts (CAFs) \cite{Hanahan:2012}. The latter make up the majority of the desmoplastic stroma and have been shown to promote the growth of primary tumors \cite{Kalluri:2016,Biffi:2021,Sahai:2020}.

Understanding how CAFs influence the architecture of collagen is key to improving our understanding of cancer cell invasion through a dense collagenous stroma as observed in breast, lung and pancreatic cancers.  At macroscopic scales, this can be studied with gel-contraction assays \cite{Bell:1979,Dallon:2008,Mikami:2016} and bead tracking \cite{Steinwachs:2015,Hall:2013,Pakshir:2019,Mark:2020}, whereby the mechanical deformation of the collagen matrix is monitored without having to explicitly consider the underlying remodelling of the collagen network. More detailed understanding of the process, based on local influence of individual cells on the matrix and the biochemical interactions involved can be obtained using atomic-force and confocal microscopy at the scale of the cells \cite{Frye:2018}. However, few studies have attempted to reconcile the macroscopic and cellular approaches to characterize the influence of cells on collagen architecture.

Developing image analysis tools to characterize the collagen-fiber network and its space- and time-dependent microstructural modifications in the context of cellular migration is challenging for a variety of reasons. In particular, the low signal-to-noise ratio of imaging techniques suitable for time-resolved studies at the required resolution precludes standard image analysis methods based on the segmentation of the fibrillar structures \cite{Bredfeldt:2014,Liu:2018}. Grey-tone image analysis methods have been developed in this context, but they focus on the analysis of fibre orientation using a variety of methods such as pixel-wised gradient estimations \cite{Boudaoud:2014}, Fourier transforms \cite{Pijanka:2019}, or mathematical morphological operations \cite{Altendorf:2012}. In addition to fibre orientation, however, key aspects of ECM remodelling concern the spatial distribution of collagen fibres. Grey-tone image analysis methods that can address that question, and statistically capture possibly non-homogeneous fibre densification patterns, are yet to be developed. 

The paper proposes such a method based on the stochastic modelling of grey-tone structures in microscopy images, and on the identification of the model parameters from their correlation functions and grey-tone histograms. Acellular collagen gels with different concentrations are considered first, in order to develop and validate the methodology. The question of ECM remodelling is then addressed in the context of a collagen-embedded CAF spheroid model, and the local evolution of the fibre network that accompanies cellular migration is studied in space- and time-resolved way.

\section{Materials and Methods}

Dulbecco's modified Eagle medium (DMEM), L-glutamine, sodium pyruvate, penicillin, streptomycin and 0.25\% trypsin/EDTA solution were purchased from ThermoFisher Scientific. Recombinant platelet-derived growth factor BB isotype (PDGF-BB) was obtained from R\&D Systems. Fetal bovine serum (FBS), sodium bicarbonate, carboxymethylcellulose and 10$\times$ concentrated DMEM, high viscosity carboxymethylcellulose sodium salt were obtained from Sigma-Aldrich. High concentration acid soluble native type I rat tail collagen was purchased from Corning and DQ-Collagen\texttrademark type I from bovine skin fluorescein conjugate was obtained from ThermoFisher. SPY650-DNA, a non-toxic, cell permeable and highly specific live cell DNA probe was purchased from Spirochrome.

\subsection{Preparation of collagen gels}

Type I collagen gels were prepared for imaging of collagen architecture and spheroid invasion assays. Collagen gels of 2.0 mg/mL and 3.0 mg/mL were prepared by diluting the stock collagen solution (8-11 mg/mL) with 10x concentrated DMEM, NaHCO$_3$, 1N NaOH, milliQ H$_2$O and neutralized to pH 7.2. To visualize collagen fibers by confocal fluorescence microscopy (CFM), fluorogenic DQ-collagen I was mixed with unlabeled diluted collagen as previously described \cite{DelAmo:2018}, obtaining a final concentration of 20 $\mu$g/mL. Collagen dilutions were performed and maintained on ice until use. Droplets (25 $\mu$L) of diluted collagens were spotted in 35mm-glass bottom $\mu$-dishes (Ibidi) and polymerized during 45 min at 19$^\circ$C, followed by a 30 min incubation at 37$^\circ$C. The polymerization temperature of 19$^\circ$C was chosen to provide matrix architecture comparable to {\it in vivo} with a network consisting of a mixture of a few thin bundles and many thick bundles  \cite{Geraldo:2012}. After completion of collagen polymerization, 1 mL of preheated DMEM supplemented with 5\% FBS was added to the dishes. Image acquisition was performed after 24 h of incubation at 37$^\circ$C with 5\% CO$_2$.

\subsection{Cell culture}

Mouse CAFs (CAFs) have been isolated from mammary gland tumors of mammary specific polyomavirus middle T antigen overexpression mouse model (MMTV-PyMT) at 12 weeks as previously described \cite{Primac:2019} and immortalized with the pLenti HPV16 E6-E7 RFP, expressing a cytoplasmic red fluorescent protein (RFP). CAFs were grown in high-glucose DMEM supplemented with 10\% FBS, 2mM L-glutamine, 1mM sodium pyruvate, 100IU/ml penicillin, 100$\mu$g/mL streptomycin. Cultures were maintained at 37$^\circ$C with 5\% CO$_2$ until their confluence reaches about 80\%.

\subsection{Spheroid invasion assay}

Spheroids were prepared by seeding 1000 CAFs in 100 $\mu$L of spheroid formation medium composed of 0.22 $\mu$m-filtered DMEM medium supplemented with 10\% FBS and 20\% carboxymethylcellulose 4000 centipoise. Cells were seeded in round-bottom non-adherent 96-well plates (CELLSTAR, Greiner Bio-One) and centrifugated at 1000 rpm for 5 min. Plates were incubated at 37$^\circ$C with 5\% CO$_2$ for 48 hours to promote spheroid formation. The content of each well was transferred in a petri dish with a 200 $\mu$L pipette (with cutted tip) and individual spheroids were collected under a binocular microscope with a 10 $\mu$L pipette. Each spheroid was resuspended in 23 $\mu$L of diluted collagen (2 mg/mL) and spotted as a 25 $\mu$l drop in a prechilled 8 well-glass bottom chamber slide (Ibidi). The slides were then transferred immediately to 19$^\circ$C and flipped to maintain the spheroids in the middle of the collagen drop (preventing their sedimentation to the glass surface or to the collagen/air interface). The extent of collagen polymerization and spheroid positioning were carefully controlled by microscopic examination throughout the polymerization step. After 30 min at 19$^\circ$C, the slides were transferred at 37$^\circ$C to complete the polymerization. Preheated culture medium (300 $\mu$L/well) supplemented with 5\% FBS, 10 ng/mL PDGF-BB and SPY650-DNA (1000-fold dilution) was added to the slides and time-lapse imaging was initiated within an hour after collagen polymerization.

\subsection{Confocal microscopy}

Images of the three-dimensional (3D) collagen gels were acquired with an inverted confocal laser scanning microscope (LSM 880 Airyscan Elyra S1, Zeiss) with a Plan-Neofluar 10$\times$/0.30 N.A. or a Plan-Neofluar 20$\times$/0.50 N.A. objective (Zeiss). The DQ-collagen containing gels were excited with 488 nm laser. Non-fluorescent collagen gels were imaged by confocal reflectance microscopy (CRM) in Airyscan high resolution mode with a simultaneous excitation of the matrix by 488 nm and 633 nm lasers. To avoid edge effects, images were acquired at least 100 $\mu$m away from the gel border, avoiding regions close to the gel/glass and gel/medium interfaces. To visualize collagen fibres of acellular gels (gels containing no cells), samples were imaged both by CFM and CRM, using the 20$\times$ objective. The resulting images have dimensions of 1000$\times$1000$\times$90 voxels with anisotropic voxel size of 0.42$\times$0.42$\times$1.10 $\mu$m$^3$, corresponding to a physical volume of approximately 500$\times$500$\times$100 $\mu$m$^3$.

Time-lapse imaging of spheroid-containing collagen gels was performed using the 10$\times$ objective with the samples incubated at 5\% CO$_2$ and 37$^\circ$C in the on-stage incubator (Okolab). The collagen matrix was imaged by CRM in Airyscan high resolution mode with a 1.4$\times$ digital zoom (scaling per pixel: 0.59 $\mu$m $\times$ 0.59 $\mu$m $\times$ 3.29 $\mu$m) and CAF were imaged by CFM in Fast Airyscan mode (scaling per pixel: 0.91$\mu$m $\times$ 0.91 $\mu$m $\times$ 3.29 $\mu$m). CAF images were rescaled to fit the images of the collagen matrix. CAF-derived RFP was excited by a 561 nm laser and detected at 591 nm. The SPY650 DNA nuclear stain was excited by a 633nm laser and the signal was captured at 654 nm. The imaging procedures for the spheroids and the matrix required a sequential imaging of the same tridimensional zone, leading to two 3D image files. Images were recorded every 30 min up to 16 h. 3D stacks were obtained at a step size of 2-$\mu$m intervals. Type I collagen fibrils have a diameter ranging from 20 nm to several hundred nm \cite{Christiansen:2000} while fibres are larger in diameter. Given that the size of each pixel is 0.42 and 0.59 $\mu$m (for 20$\times$ and 10$\times$ objectives, respectively), it is not possible to distinguish fibrils from fibres \cite{Lai:2012}, therefore the term ``fibres'' was used to include both fibrils and fibres.

The raw images were converted using the Airyscan algorithm from the Zeiss Zen Black software. The images were then subjected to a histogram stretching and converted to tiff format through the Zeiss Zen Blue software. This operation turned the original 16-bit images into 8-bit images, which were corrected by histogram adjustment to maximize the conservation of valuable information.

\section{Results}

\subsection{Covariance and grey-tone correlation function}

\begin{figure}
\centering
\includegraphics[width=6cm]{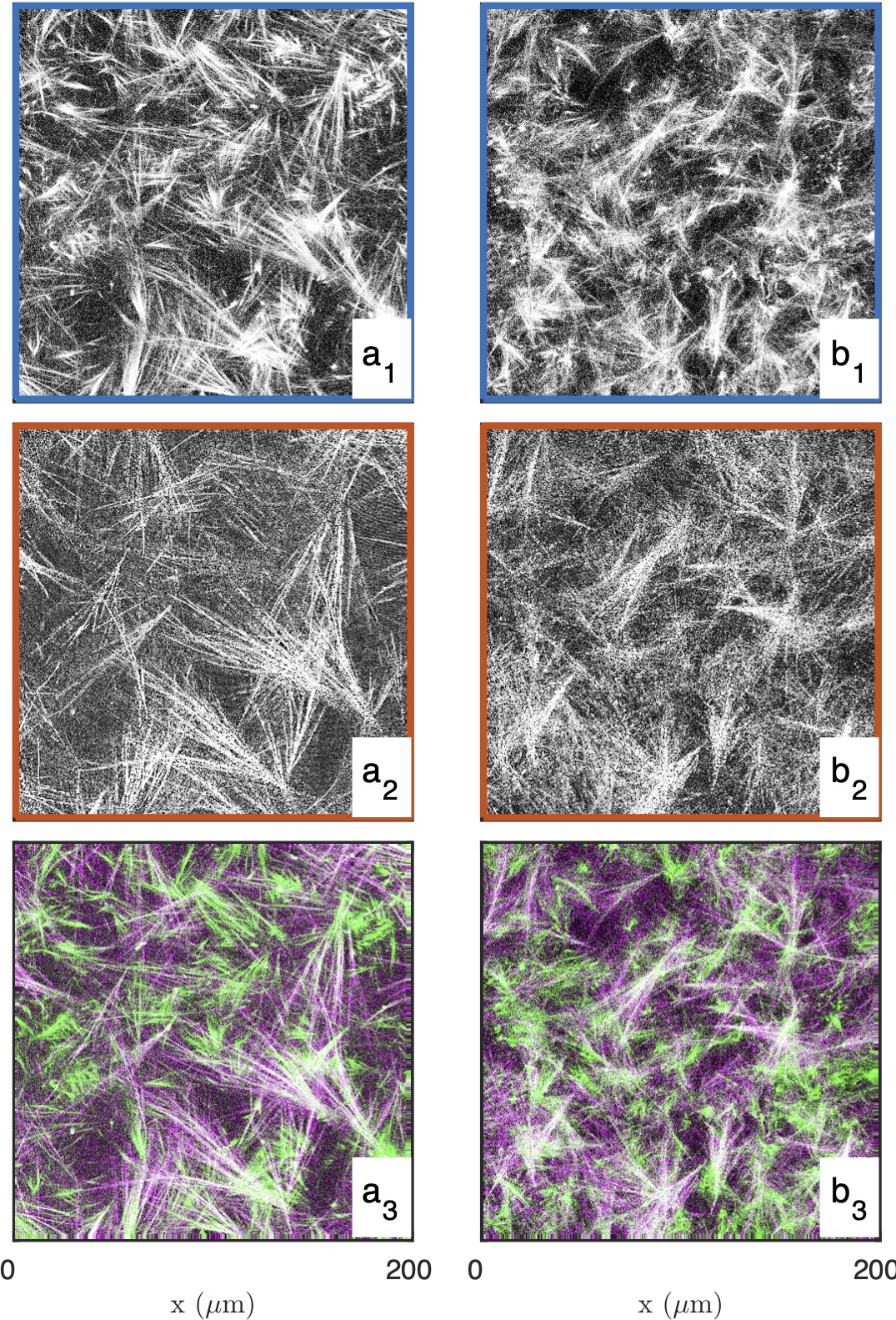}
\caption{Images of the 2 mg/mL (a) and 3 mg/mL (b) acellular gels obtained by confocal microscopy in fluorescence (a$_1$ and b$_1$) and reflectance (a$_2$ and b$_2$) modes. The bottom row displays fused images with fluorescence and reflectance data in green and magenta, respectively.}
\label{fig:gel_images}
\end{figure}

Examples of images of the dilute (2 mg/mL) and concentrated (3 mg/mL) gels obtained through confocal fluorescence (CFM) and confocal reflectance (CRM) microscopy are given in Fig \ref{fig:gel_images}. These are 2D single-$z$ images taken out of the 3D images. The structural analysis is based on 15 images such as presented in the figure, for each gel and each imaging mode. Type I collagen fibrils have a diameter ranging from 20 nm to several hundred nm \cite{Christiansen:2000} while fibres are larger in diameter. Given that the size of each pixel is 0.42 and 0.59 $\mu$m (for 20x and 10x objectives, respectively), it is not possible to distinguish fibrils and fibres \cite{Lai:2012}, therefore the term ``fibres'' was used to include both fibrils and fibres. The most salient structures are the fibre aggregates that are a few tens of micrometers across. The fused images in Fig \ref{fig:gel_images}a$_3$ and b$_3$ show that there is limited overlap between the CRM and CFM microscopy data: the fibres and aggregates are well captured in CFM but it is mostly the aggregates that are visible in the CRM data.

\begin{figure}
\centering
\includegraphics[width=8cm]{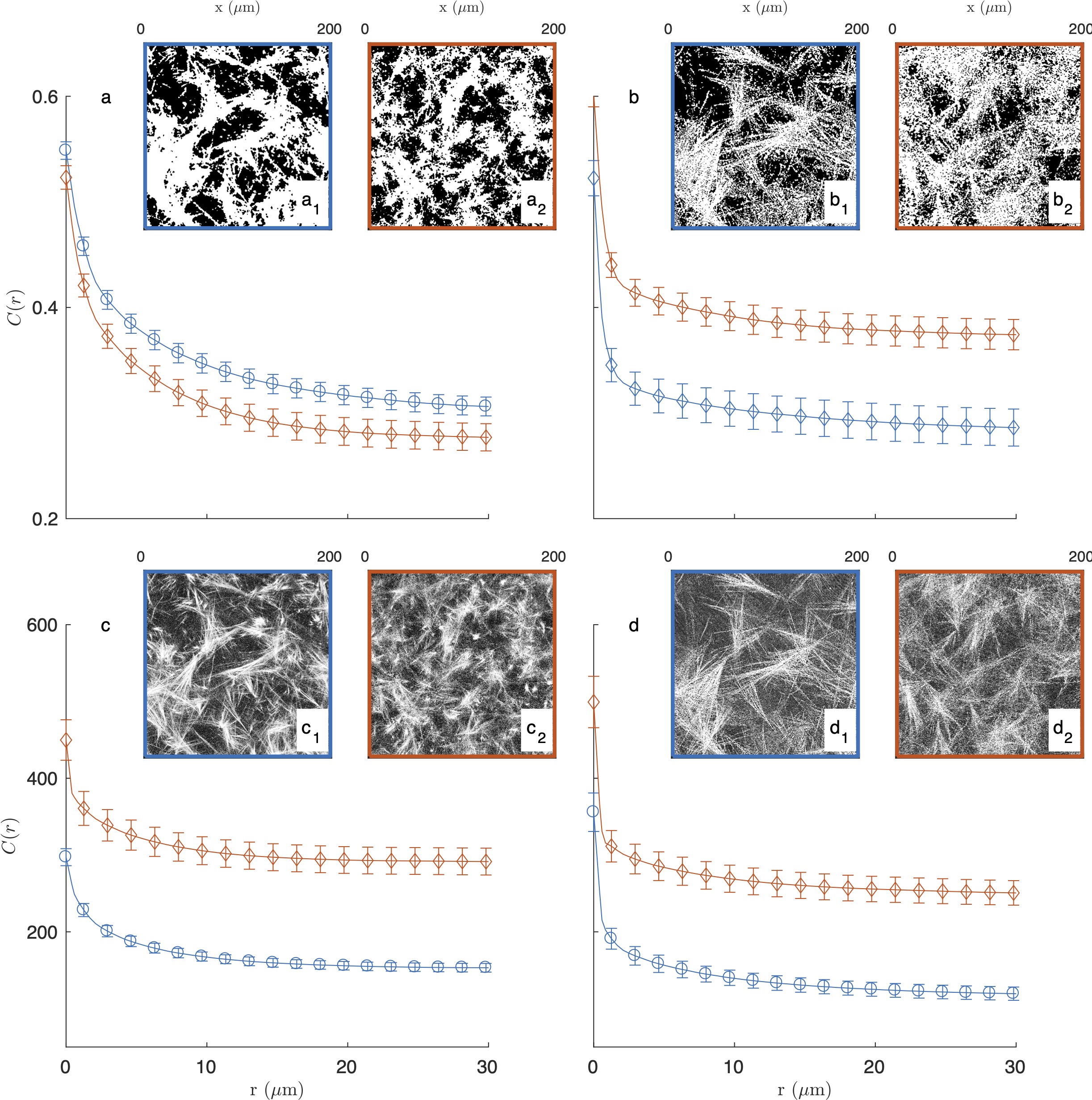}
\caption{Experimental covariance (top) and grey-tone correlation function (bottom) of the 2 mg/mL (blue circle, a$_1$, b$_1$, c$_1$, d$_1$) and 3 mg/mL (red diamond, a$_2$, b$_2$, c$_2$, d$_2$) gels, imaged in fluorescence (a, c) and reflectance (b, d) modes. The insets display one of the fifteen images used for each condition, and the error bars are the standard errors of the mean.}
\label{fig:gel_covariance}
\end{figure}

In order to follow a standard image analysis procedure, and justify further less-standard developments, we first explored a method based on image segmentation. In that spirit, the grey-tone images of the gels are converted to binary images following a method described in the Supplementary Material (see Fig. S1). Examples of segmented images of the gels, with collagen in white and the rest in black, are provided in the insets of Figs. \ref{fig:gel_covariance}a and \ref{fig:gel_covariance}b.

The segmented gel images display complex and disordered structures, which are also corrupted by noise in the case of the reflectance images (Figs. \ref{fig:gel_covariance}b$_1$ and \ref{fig:gel_covariance}b$_2$). In this context, the covariance - which describes the spatial correlation between all pixels intensities in the images \cite{Serra:1982,Lantuejoul:2002,Torquato:2002,Jeulin:2021}- was used to quantitatively analyze the gel structures. The covariances shown in Fig. \ref{fig:gel_covariance}a and \ref{fig:gel_covariance}b were obtained from fifteen images taken in the same gel with dimension 200 $\times$ 200 $\mu m^2$ as illustrated in the insets, via a Fourier-transform algorithm, and the error bars are the standard errors of the means.

The covariance $C(r)$ of say, the white component of an image, has a clear geometrical interpretation as it is defined as the probability that any couple of points at distance $r$ from one another both belong to that phase. For very small values of $r$, this definition coincides with the probability for a single point to belong to the white phase, which is numerically equal to its density $\phi_1$ normalized between 0 and 1. In the opposite limit, that is for very large distances $r$ the covariance converges to the value $\phi_1^2$, which corresponds to a horizontal asymptote in Fig \ref{fig:gel_covariance}. The shape of the covariance curve between those two limits characterizes the structures present in the images. In particular the progressive decrease of $C(r)$ over distances of a few tens of microns testifies to the presence of structures with those dimensions, which we qualitatively referred to earlier as fibre aggregates.  

The covariance of the fluorescence images of the gels (in Fig. \ref{fig:gel_covariance}a) highlights at once the unsuitability of image segmentation in the present context. The covariance of the 2mg/mL gel is found to be larger than that of the 3mg/mL gel for any $r$, which means that densities of the segmented images contradict the actual collagen concentrations of the gels. This results from the fact that collagen-rich areas of the images display a variety of grey-tones related to the local fibre density (see Fig. S1b), which information is lost during the all-or-nothing segmentation procedure. This general observation calls for grey-tone image analysis methods that preserve the structural information in the images. In that spirit, the grey-tone correlation functions $C(r)$ of the gels are shown in Fig. \ref{fig:gel_covariance}c and \ref{fig:gel_covariance}d. The latter are measured directly on the unprocessed images and they characterise the statistical correlation between grey-tones of all pixels that are at distance $r$ from one another. In the case where the images contain only the values $0$ and $1$, this definition is mathematically equivalent to the covariance. We postpone to a later section the discussion of the structural significance of the grey-tone correlation function, but we already notice at this stage that the grey-tone data in Figs. \ref{fig:gel_covariance}c and \ref{fig:gel_covariance}d scale with the actual collagen concentration of the 2mg/mL and 3mg/mL gels as they should.

\subsection{Structural models}

Covariance and grey-tone correlation functions convey indirect yet very rich structural information \cite{Aubert:2000,Jiao:2009,Gommes:2012}, which can notably be retrieved using structural models.
We here present two models aimed at extracting structural information from the covariance data in Fig. \ref{fig:gel_covariance}, which we generalize later to grey-tone correlation functions. In order to cope with the disordered structure of the gel, the models have to be stochastic \cite{Jeulin:2000,Torquato:2002,Gommes:2018,Jeulin:2021}.  

When using stochastic models, a structure is defined through probabilistic rules and this calls for specific concepts. In particular, it is convenient to introduce the indicator function of the structure $\mathcal{I}(\mathbf{x})$, which takes the value 1 if the point $\mathbf{x}$ belongs to the structure and 0 otherwise \cite{Torquato:2002}. With such definition, the density of the model is calculated as
\begin{equation} \label{eq:volume_fraction}
\phi_1 = \langle  \mathcal{I}(\mathbf{x}) \rangle
\end{equation}
where the brackets $\langle \rangle$ stand for the average value, calculated either over $\mathbf{x}$ or over all the possible realization of the model. Similarly, the covariance is calculated as the following two-point average
\begin{equation} \label{eq:covariance}
C_{11}(r) = \langle  \mathcal{I}(\mathbf{x}) \mathcal{I}(\mathbf{x} + \mathbf{r}) \rangle
\end{equation}
because the product $\mathcal{I}(\mathbf{x}) \mathcal{I}(\mathbf{x} + \mathbf{r})$ is equal to one, only if the points $\mathbf{x}$ and $\mathbf{x}+\mathbf{r}$ belong to the structure. In the latter equation, we have assumed statistical isotropy so that the dependence is only through the modulus $r = |\mathbf{r}|$. 

\subsubsection{Homogeneous fibre model}

\begin{figure}
\centering
\includegraphics[width=7cm]{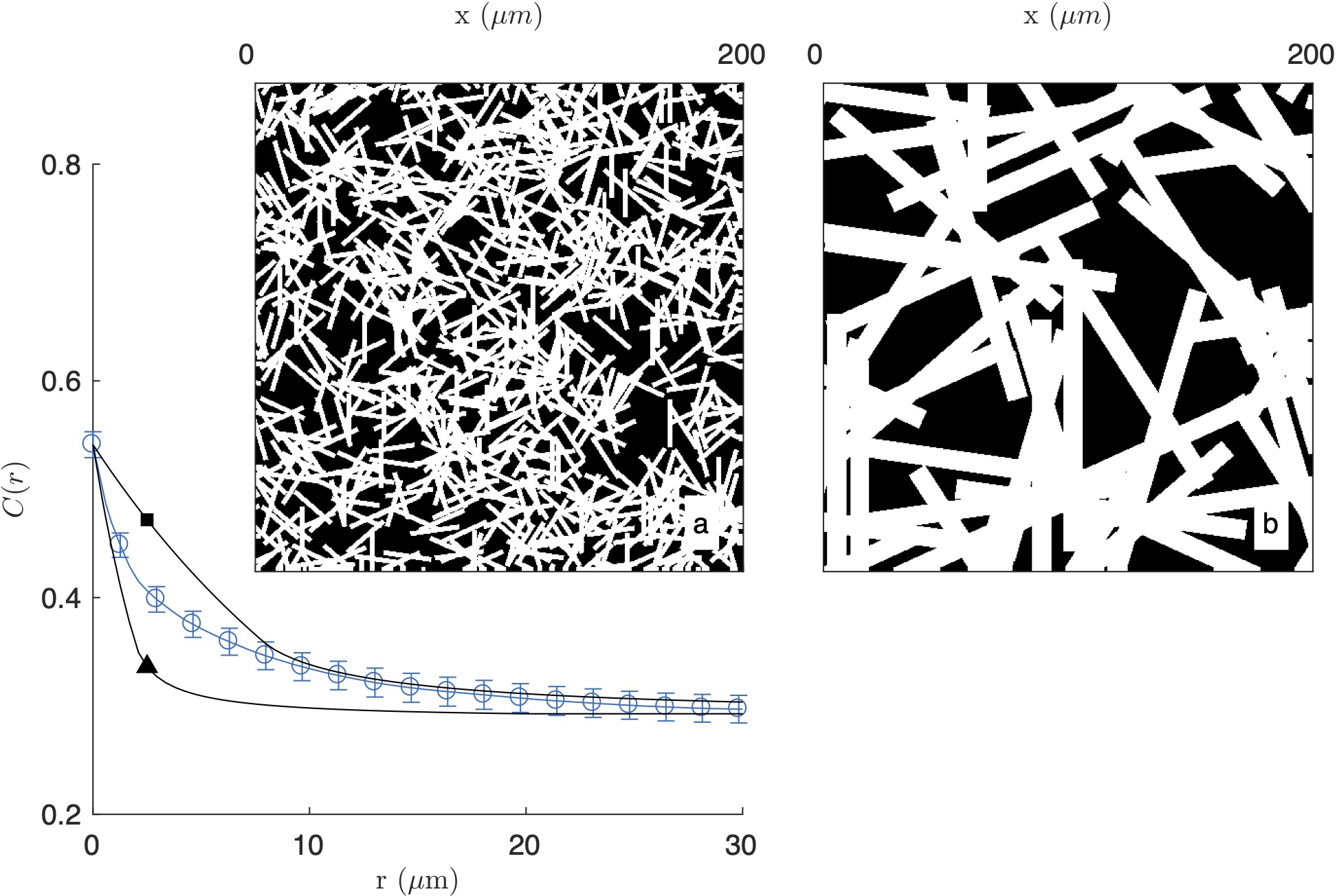}
\caption{Homogeneous fibre model, obtained as a Boolean model of randomly oriented rectangles. The two realisations are obtained with length and diameter $L_F=20$ $\mu$m, $D_F=1$ $\mu$m (a and $\blacktriangle$) and $L_F=100$ $\mu$m, $D_F=10$ $\mu$m (b and $\blacksquare$), and the calculated covariances are compared with that of the 2mg/mL gel imaged by CFM (same as Fig. \ref{fig:gel_covariance}a1)}
\label{fig:fibre_model}
\end{figure}

The simplest model we consider to analyze the structures in Fig. \ref{fig:gel_images} assumes that the gel matrix is statistically homogeneous. The model consists in tossing fibres (modelled as elongated rectangles) with random position and orientation, as sketched in Fig. \ref{fig:fibre_model}. Such model is described by three parameters, namely: the number density of fibres $\theta$ (unit $\mu$m$^{-2}$), as well as their length $L_F$ and diameter $D_F$ (both in units of $\mu$m). Note that the statistical homogeneity of the model does not rule out the existence of aggregates, which form when a large number of fibres coincidentally fall in the same region of space. The same holds for pores in the gel matrix. 

The homogeneous fibre model is a particular case of a Boolean model \cite{Serra:1982,Jeulin:2021}, for which the density and covariance are known analytically. In particular, the fibre density is
\begin{equation} \label{eq:Boolean_phi}
\phi_F = 1 - \exp\left[ - \eta \right]
\end{equation}
where $\eta = \theta D_F L_F$ is the density one would expect in absence of fibre overlap. The covariance is given by the following expression
\begin{equation} \label{eq:Boolean_C}
C_{FF}(r) = 2 \phi_F - 1 + (1-\phi_F)^2  \exp\left[ \eta K_F(r) \right]
\end{equation}
where $K_F(r)$ is the geometrical covariogram of the randomly-oriented fibres, which is calculated as
\begin{equation} \label{eq:KF}
K_F(r) = \frac{2}{\pi} \int_0^{\pi/2} \left[1 - \frac{r}{D_F} \sin(t) \right] \left[1 - \frac{r}{L_F} \cos(t) \right] \textrm{d}t
\end{equation}
for $r<D_F$ and the upper integration bound is replaced by $\textrm{asin}(D_F/r)$ for $r \ge D_F$. 

The covariance of the homogeneous fibre model is plotted in Fig. \ref{fig:fibre_model} for two fibres sizes. For the purpose of illustration, the model is compared with the experimental covariance of the segmented image of the 2mg/mL gel (from CFM). In the figure, the number of fibres $\theta$ is chosen to achieve a density $\phi_F \simeq 0.54$ comparable to the segmented image. As a consequence, the asymptotic values of $C_{FF}(r)$ are a close match to the gel covariance both for $r = 0$ and for $r \to \infty$. The shape of the covariance at intermediate distances, however, cannot be captured by the homogeneous model. The model can account for either the small- or large-$r$ data but not the two simultaneously. In the former case, the model captures the small-scale structure of the gel (Fig. \ref{fig:fibre_model}a) but it is unable to account for the aggregates, which are found to be more prevalent than what can be expected from randomness alone. In the latter case, the large-scale structure of the gel is reasonably reproduced, but this is done by replacing fibre aggregates by unrealistically large rectangles (Fig. \ref{fig:fibre_model}b).

\subsubsection{Fibre aggregates model}

The inability of the homogeneous fibre model to reproduce the experimental covariance of gels proves that structures larger than individual fibres are more frequent than what can be accounted for by statistical fluctuations alone. The existence of such large pores and aggregates is notably apparent when comparing the realization of the homogeneous model in Fig. \ref{fig:fibre_model}a with the insets of Fig. \ref{fig:gel_covariance}.

To address this issue, we introduce a second model that builds on two distinctly different structures. At the smallest scale the structure is assumed to be that of homogeneously-distributed fibres, corresponding to indicator function $\mathcal{I}_F(\mathbf{x})$.The larger-scale structure, however, is accounted for by creating the indicator function of the entire structure $\mathcal{I}(\mathbf{x})$ through the following multiplication
\begin{equation} \label{eq:Indicator_product}
\mathcal{I}(\mathbf{x}) = \mathcal{I}_F(\mathbf{x}) \times \mathcal{I}_A(\mathbf{x}) 
\end{equation}
where $\mathcal{I}_A(\mathbf{x})$ is the indicator function of the aggregates, equal to one if $\mathbf{x}$ is inside an aggregate. Mathematically, Eq. (\ref{eq:Indicator_product}) is equivalent to starting with a homogeneous fibre structure and subsequently carving pores out of it, using $\mathcal{I}_A(\mathbf{x})$ as a mathematical cookie-cutter.

Independently of the specific models chosen for $\mathcal{I}_F(\mathbf{x})$ and $\mathcal{I}_A(\mathbf{x})$, evaluating the average value of Eq. (\ref{eq:Indicator_product}) yields the following density for the two-scale structure
\begin{equation} \label{eq:fraction_product}
\phi_1 = \phi_F \phi_A
\end{equation}
where $\phi_A$ is the density of the aggregates, and $\phi_F$ is the density of the fibres within the aggregates. Equation (\ref{eq:fraction_product}) results from the general definition of the density in Eq. (\ref{eq:volume_fraction}), with the assumption that the fibre and aggregate models are statistically independent from one another. The same assumption provides the following relation for the covariance of the solid phase 
\begin{equation}
C_{11}(r)= C_{FF}(r) C_{AA}(r) 
\end{equation}
as a consequence of Eq. (\ref{eq:covariance}). 

For the small-scale structure, we assume the same fibre model as considered earlier, with density $\phi_F$ and covariance $C_{FF}(r)$ given in Eqs. (\ref{eq:Boolean_phi}) and (\ref{eq:Boolean_C}). As the aggregates are very disordered with no well-defined shape, we model them with a clipped Gaussian-field approach \cite{Quiblier:1984,Berk:1987,Gommes:2018} as described in the Supplementary Material. The two parameters of the large-scale model are the density of the aggregates $\phi_A$ and a single characteristic length $L_A$ that controls their size. 

\begin{figure}[]
\centering
\includegraphics[width=8cm]{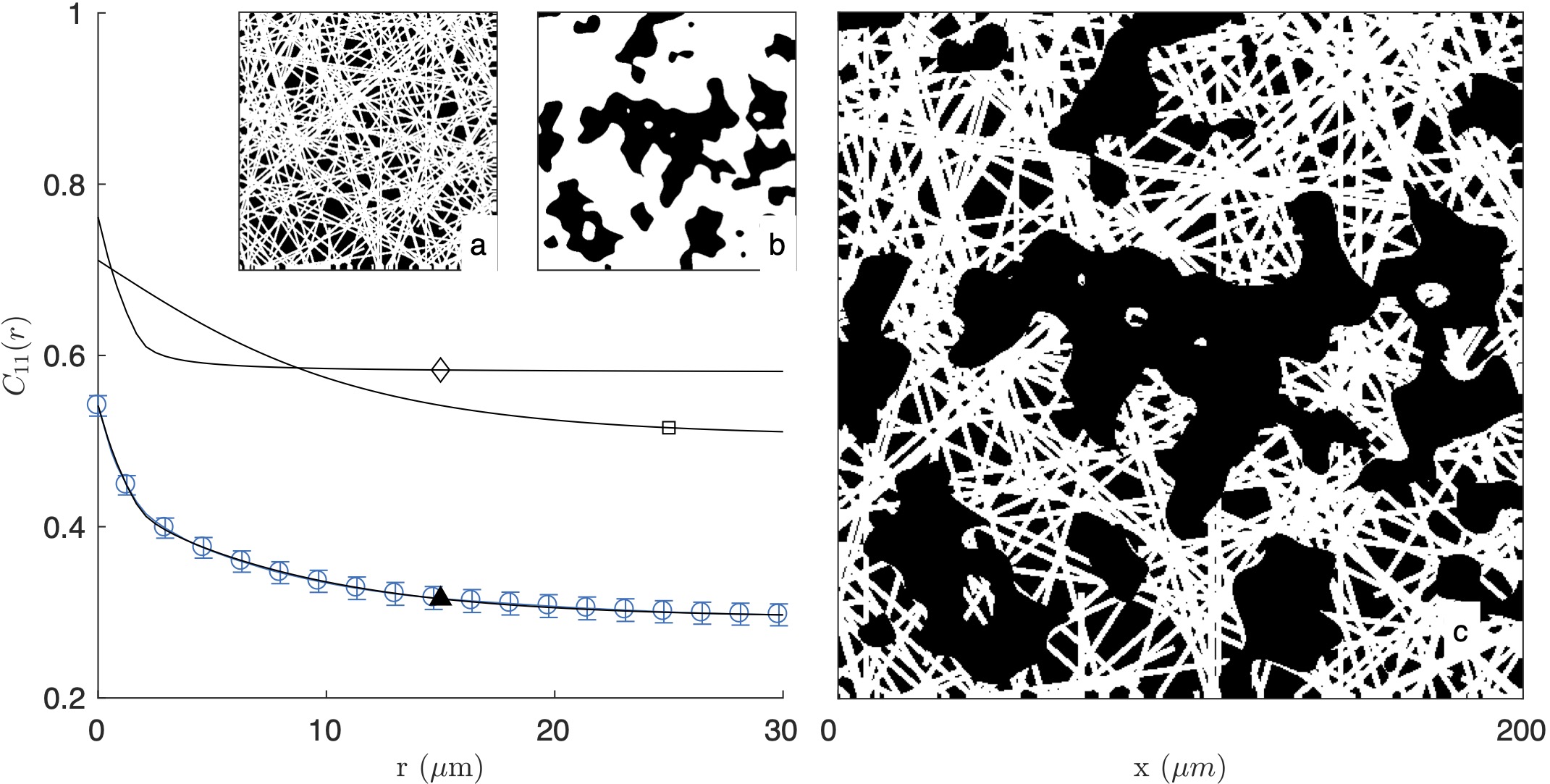}
\caption{Covariance of the fibre-aggregate model, with homogeneous fibre model at small scale (a and $\diamond$) and clipped Gaussian-field model for the large-scale aggregates (b and $\square$), combined to yield a two-scale structure (c and $\blacktriangle$). The calculated covariance is compared with that of the 2mg/mL gel imaged by fluorescence microscopy (same as Fig. \ref{fig:gel_covariance}a1)}
\label{fig:model_aggregates}
\end{figure}

As illustrated in Fig. \ref{fig:model_aggregates} for the segmented image of the 2mg/mL gel in fluorescence mode, the fibre-aggregate model captures well the experimental covariance of the gels. For the fitting of the data, the parameter $L_F$ in the fibre model is irrelevant as the actual length of the fibres is controlled by the size of the aggregates. Equation \ref{eq:KF} was therefore simplified to its limit $L_F \to \infty$. The values of the remaining parameters are gathered in Tab. \ref{tab:fitted_parameters} for the two gels and the two imaging modes.  From the fitted parameters $\phi_F$ and $\phi_A$ of the fibre-aggregate model, the total density of the fibres $\phi_1$ was calculated through Eq. (\ref{eq:fraction_product}) and is also reported in Tab. \ref{tab:fitted_parameters}.

\begin{table}
\centering
\caption{Structural parameters of the gels, obtained from fitting the covariance (binary images) or the correlation function and histogram (grey-tone images) with the fibre-aggregate model.}
\begin{tabular}{lllcccc|c}
\toprule
Imaging  & Processing &  Gel & $\phi_F$ (-) & $\phi_A$ (-) & $D_F$ ($\mu$m) & $L_A$ ($\mu$m) & $\phi_1$ (-)\cr
\hline
\midrule
\multirow{4}{*}{CFM} & \multirow{2}{*}{Binary} & 2 mg/mL &  $0.76 \pm 0.03$  & $0.72 \pm 0.01$ & $1.9 \pm 0.2$  & $8.2 \pm 1$  & $0.55 \pm 0.01$   \cr
 &  & 3 mg/mL &  $0.76 \pm 0.01 $  & $0.69 \pm 0.01 $ & $1.5 \pm 0.1$ & $6.3 \pm 0.4$ & $0.52 \pm 0.01$   \cr
\cmidrule{3-8}    
  & \multirow{2}{*}{Grey-tone}  & 2 mg/mL &  $0.73 \pm 0.04$  & $0.32 \pm 0.05$ & $1.9 \pm 0.1$  & $6.6 \pm 1$  & $0.23 \pm 0.03$  \cr
  &  &  3 mg/mL &  $0.78 \pm 0.06$  & $0.49 \pm 0.04$ & $2.3 \pm 0.1$ & $7.7 \pm 1$  & $0.38 \pm 0.06$  \cr                    
 \midrule
 \hline
\multirow{4}{*}{CRM} & \multirow{2}{*}{Binary} &  2 mg/mL &  $0.60 \pm 0.02$ & $0.88 \pm 0.04$ & $0.7 \pm 0.09$  & $15 \pm 1$ & $0.53 \pm 0.04$   \cr 
 &  &  3 mg/mL &  $0.67 \pm 0.01$  & $0.90 \pm 0.02$ & $0.90 \pm 0.03$ & $13 \pm 1$  & $0.61 \pm 0.01$ \cr 
\cmidrule{3-8}
  & \multirow{2}{*}{Grey-tone}  &  2 mg/mL &  $0.78 \pm 0.05$ & $0.41 \pm 0.03$ & $0.98 \pm 0.2$  & $7.6 \pm 0.5$ & $0.32 \pm 0.01$   \cr 
 &  &  3 mg/mL &  $0.61 \pm 0.06$ & $0.56 \pm 0.01$ & $0.62 \pm 0.1$  & $14 \pm 7$ & $0.34 \pm 0.02$   \cr 
\bottomrule
\hline
\end{tabular}
\begin{flushleft} $\phi_F$, $D_F$: density and diameter of the fibres; $\phi_A$, $L_A$: density and size of the aggregates; $\phi_1$: total fibre density. The error bars are the standard deviations observed from the fits of the fifteen images in each condition.
\end{flushleft}
\label{tab:fitted_parameters}
\end{table}

\begin{figure}
\centering
\includegraphics[width=6cm]{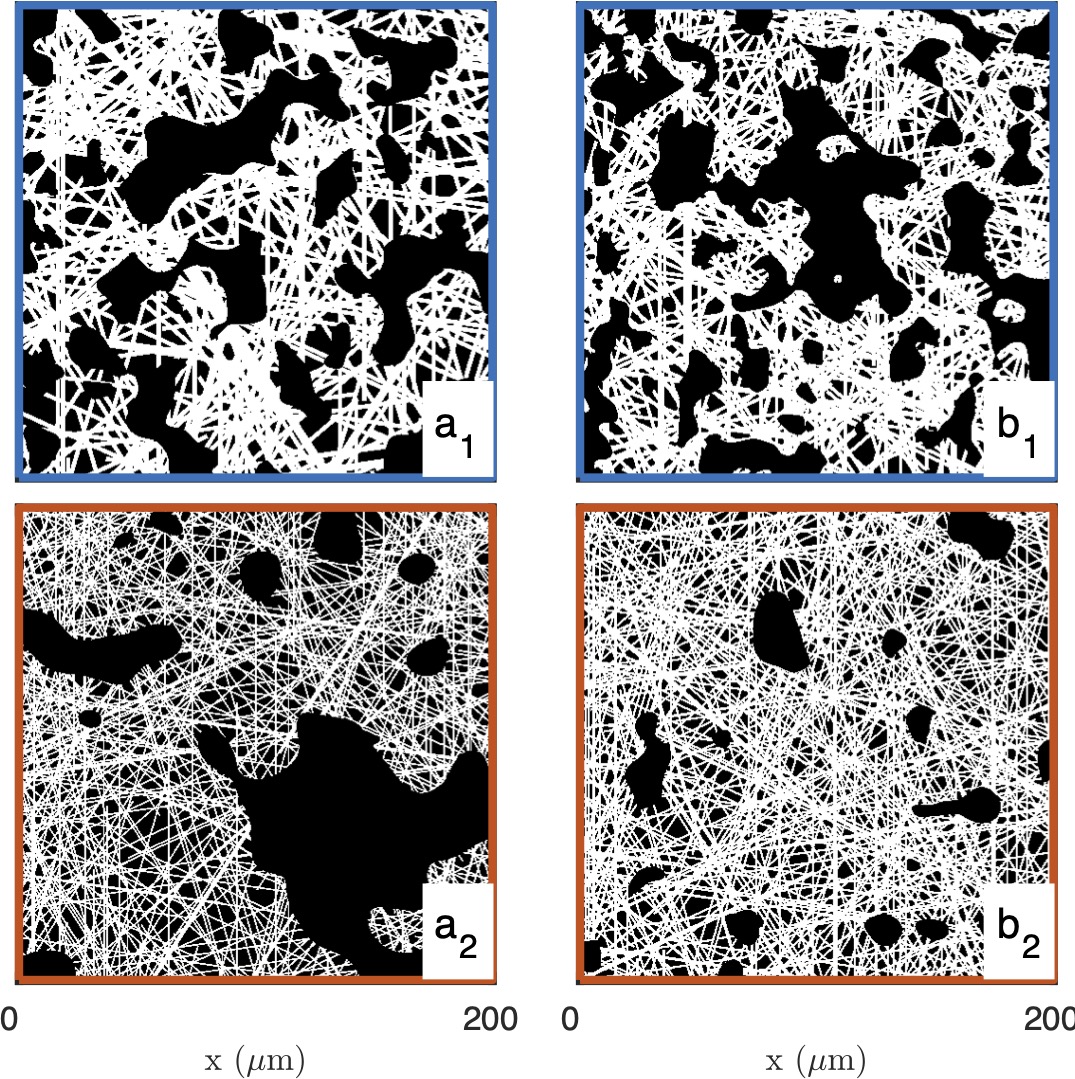}
\caption{Realizations of the fibre aggregate model, for the 2mg/mL (a$_1$, a$_2$) and 3mg/mL (b$_1$, b$_2$) gels, based on the average values of the parameters fitted from the covariance of the CFM (a$_1$ and b$_1$) and CRM (a$_2$ and b$_2$) segmented images.}
\label{fig:realization_aggregates}
\end{figure}

Realizations of the fibre aggregate model for the two gels and the two imaging modes are given in Fig. \ref{fig:realization_aggregates}. These realizations illustrate the structures captured by the covariance in the segmented images of the gels, and they largely illustrate the drawbacks of a data-analysis based on image segmentation. As mentioned earlier, the segmentation of the fluorescence data leads to inconsistencies between the density of the segmented images and the gel concentrations. In addition, the significant noise in the reflectance data lead to unrealistically large values for the aggregate density (see Figs. \ref{fig:gel_covariance}b$_1$-b$_2$ and \ref{fig:realization_aggregates}b$_1$-b$_2$).

\subsubsection{Greytone model}

In order to overcome the limitations of image segmentation, we expand here the binary fibre-aggregate model so as to allow the identification of its parameters directly from grey-tone images. The binary models ignore the fact that large image intensities are associated with large local fibre concentrations. In the grey-tone model, we therefore assume that every fibre contributes additively (by a quantity $\Delta$) to the local grey-tone of the image. Namely, the image intensity where 2, 3 or more fibres overlap is $2 \Delta$, $3 \Delta$, etc. The model also accounts explicitly for the noisiness of the data, through uncorrelated Gaussian noise $n(\mathbf{x})$ with variance $\sigma^2$, and for a background intensity $b$. 

These assumptions can be  formally written by expressing the local image intensity at point $\mathbf{x}$ as
\begin{equation} \label{eq:greytone}
I(\mathbf{x}) = b + \mathcal{I}_A(\mathbf{x}) \times \Delta \sum_i  \mathcal{I}_F^{(i)}(\mathbf{x}) + n(\mathbf{x})
\end{equation}
where $\mathcal{I}_A(\mathbf{x})$ is the indicator function of the aggregates as before, and the sum accounts for the overlapping of the fibres. In the sum, each term $\mathcal{I}_F^{(i)}(\mathbf{x})$ is the indicator function of a homogeneous fibre model with vanishingly small number density $\eta^{(i)}$ so as to the overlapping of fibres. The fibre density $\phi_F$ is then obtained again through Eq. (\ref{eq:Boolean_phi})., with the total number density $\eta = \sum_i \eta^{(i)}$. 

The statistical independence of all contributions in Eq. (\ref{eq:greytone}) -namely, $\mathcal{I}_A(\mathbf{x})$, $n(\mathbf{x})$ and $\mathcal{I}_F^{(i)}(\mathbf{x})$ for all $i$'s- enables one to calculate the corresponding grey-tone correlation function. This is done through to application of Eq. (\ref{eq:covariance}) to the image intensity in Eq. (\ref{eq:greytone}), and the result is
\begin{equation} \label{eq:greytone_correlation}
C(r) = \langle I \rangle^2 + \eta \Delta^2 K_F(r) C_{AA}(r) + (\eta \Delta)^2 \left( C_{AA}(r) - \phi_A^2 \right) + \sigma^2 \delta(r)
\end{equation}
where $K_F(r)$ is the geometrical covariogram of the fibres, $C_{AA}(r)$ is the covariance of the aggregates, the last term is the correlation function of uncorrelated Gaussian noise where $\delta(r)$ is equal to 1 if $r=0$ and to 0 otherwise, and
\begin{equation}
\langle I \rangle = b + \eta \phi_A \Delta 
\end{equation}
is the average intensity of the image.

\begin{figure}
\centering
\includegraphics[height=7cm]{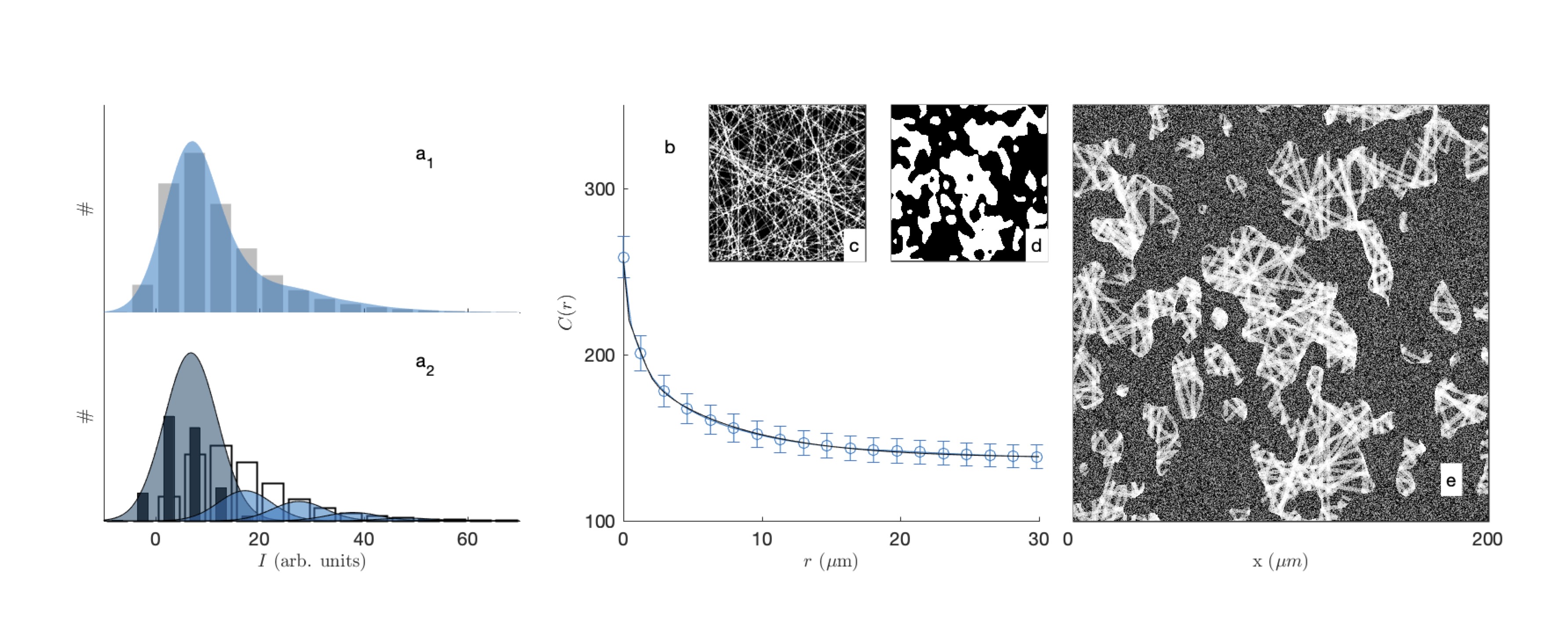}
\caption{Greytone fibre-aggregates model of the 2mg/mL type-I collagen gel, imaged by fluorescence microscopy, with (a1) histogram of grey levels (bars: experimental values; blue shade: model), (b) the correlation function (dots: experimental value; solid line: model), as well as realizations of the fibres (c), aggregates (d) and complete model (e). In (a$_2$) the values inferred from the segmented images in the pores (back bars) and fibre areas (white bars) are compared with the model contributions to the background (dark shade) and increasing number of overlapping fibres (bright shades). }
\label{fig:model_grey}
\end{figure}

In addition to correlation functions, the grey-tone modelling enables one to extract structural information from the grey-tone distribution itself. Because the fibres in the model are distributed according to a Boolean process, the probability for exactly $k$ fibres to overlap at any point of space is a Poisson variable, namely
\begin{equation} \label{eq:Poisson}
\textrm{Prob} \Big\{ k \textrm{ overlapping fibres} \Big\} = \frac{\eta^k \exp[-\eta]}{k !}
\end{equation}
which contains Eq. (\ref{eq:Boolean_phi}) as a particular case, when all probabilities for $k$ larger or equal to 1 are added. As each fibre contributes a quantity $\Delta$ to the local intensity, Eq. (\ref{eq:Poisson}) can also be interpreted as the probability for observing the intensity $k \times \Delta$ for any point inside an aggregate. Accounting also for the background intensity $b$ and for the Gaussian noise, the intensity distribution is therefore
\begin{equation} \label{eq:greytone_histogram}
f(I) = (1 - \phi_A \phi_F) g_\sigma[I-b] + \phi_A \exp[-\eta] \sum_{k=1}^\infty \frac{\eta^k}{k !} g_\sigma[I-(b+k\Delta)]
\end{equation}
where 
\begin{equation}
g_\sigma[x] = \frac{1}{\sqrt{2 \pi} \sigma} \exp[-\frac{x^2}{2 \sigma^2} ]
\end{equation}
is the centred Gaussian probability density. The quantity $f(I) \textrm{d}I$ is the probability for a pixel to have intensity in the interval $[I, I + \textrm{d}I]$. In Eq. (\ref{eq:greytone_histogram}) the first term accounts for grey-tone distribution of the background, outside the aggregates or between fibres within the aggregates. In the second term, the sum over $k$ is on the increasing number of overlapping fibres within the aggregates.

\begin{figure}
\centering
\includegraphics[width=7cm]{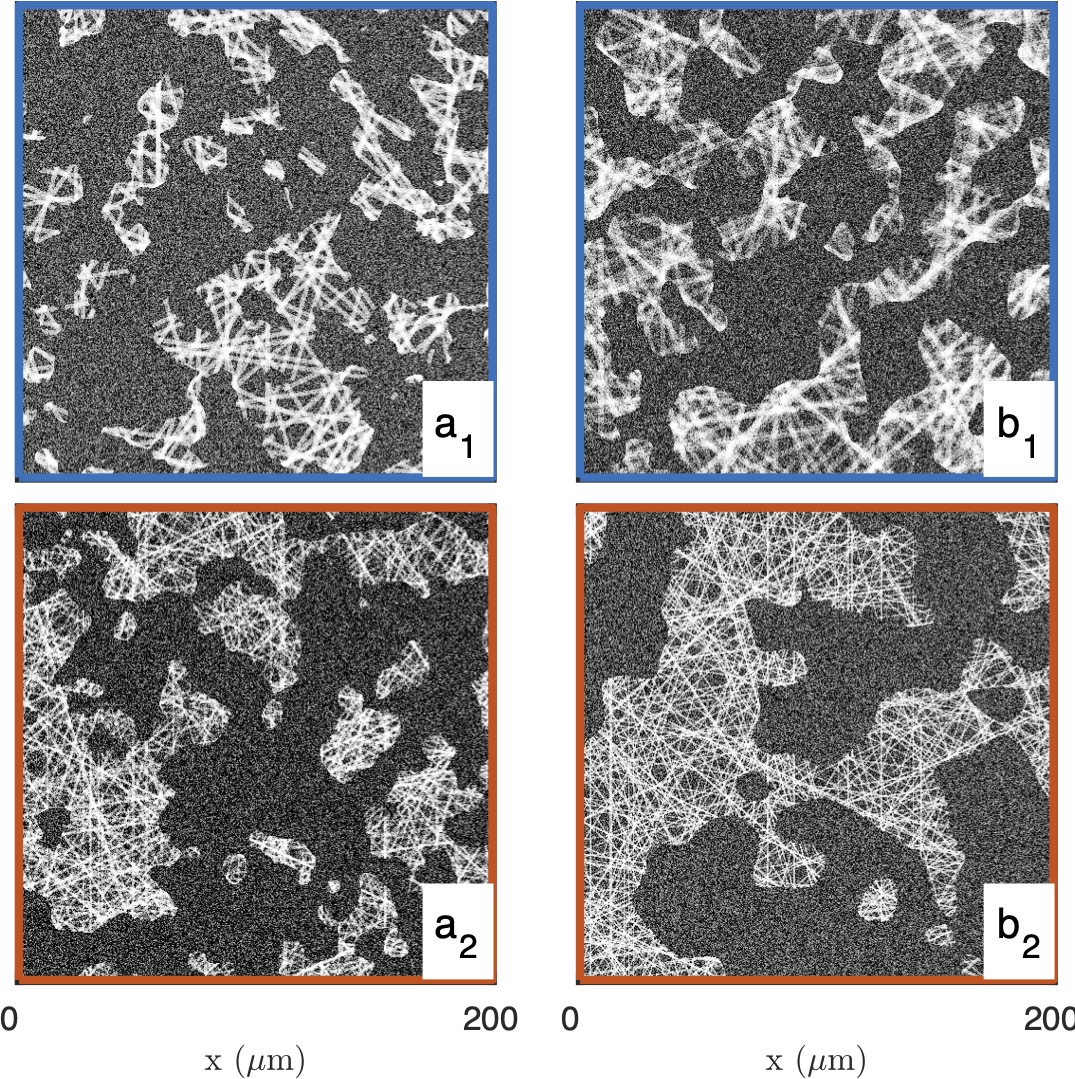}
\caption{Realizations of the grey-level fibre aggregate model, for the 2mg/mL (a$_1$, a$_2$) and 3mg/mL (b$_1$, b$_2$) gels, based on the average values of the parameters fitted from the fluorescence (a$_1$ and b$_1$) and reflectance (a$_2$ and b$_2$) images.}
\label{fig:realizations_grey}
\end{figure}

Figure \ref{fig:model_grey} illustrates the fitting of the microscopy images with the grey-tone model. The model captures nicely both the distribution of grey levels (Fig. \ref{fig:model_grey}a1) and the correlation function (Fig. \ref{fig:model_grey}b). As an independent check, Fig. \ref{fig:model_grey}a$_2$ compares the modelled distribution of grey tones with the values measured in the regions of the image that were classified as pores or fibres in the segmented images (see Fig. S1). The comparison should be considered with caution, as it is the unsuitability of the segmentation that justified the present grey-tone approach. The agreement of the two methods, however, seem reasonable. The various contributions to the fibre intensity in Fig. \ref{fig:model_grey}a$_2$ also testify to the importance of fibre overlap for the image analysis. The values of the fitted parameters on the two gels and two imaging modes are reported in Tab. \ref{tab:fitted_parameters}. Realizations obtained from the average values of the fitted parameters are given in Fig. \ref{fig:realizations_grey}. 

\section{Time- and space-resolved analysis of the gel structure in the spheroid model}

\begin{figure}[b!]
\centering
\includegraphics[width=13cm]{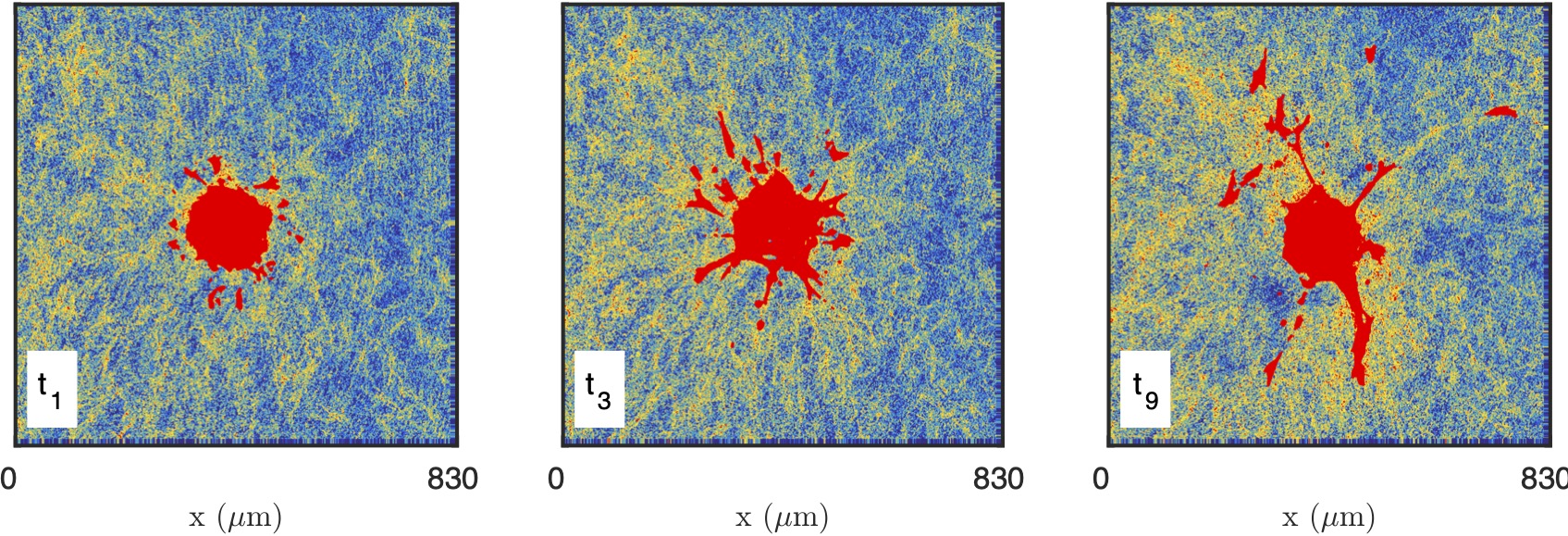}
\caption{Images of the CAF spheroid and surrounding gel at various times: $t_1=30$ min, $t_3=210$ min, $t_9$=750 min. The images correspond to one height $z$ cutting through the spheroid, with the cells shown in red and the density of the gel coded from blue to yellow.}
\label{fig:spheroid_images}
\end{figure}

Time-resolved reflectance images of the platelet-derived growth factor-BB (PDGF-BB)-treated CAF spheroid and surrounding gel are shown in Fig. \ref{fig:spheroid_images}, for times ranging from 30 min to 750 min. In order to characterize the spatiotemporal dynamics of cell interaction with the surrounding collagen matrix during cell migration, the system was imaged at lower resolution than the acellular gels considered so far. At the scale of the spheroid, individual fibres are not visible and the only structures detected in the gels are the aggregates. The images testify to a progressive local densification of the gel structure close to the spheroid and migrating cells but it is unclear whether other structural characteristics of the gel evolve as well. Moreover, as the gel heterogeneity seems to develop at the same scale as the spheroid, it is difficult to ascertain whether the effect of the cells is through direct contact with the gel, or if long-range effects are also operative.

\begin{figure}
\centering
\includegraphics[width=15cm]{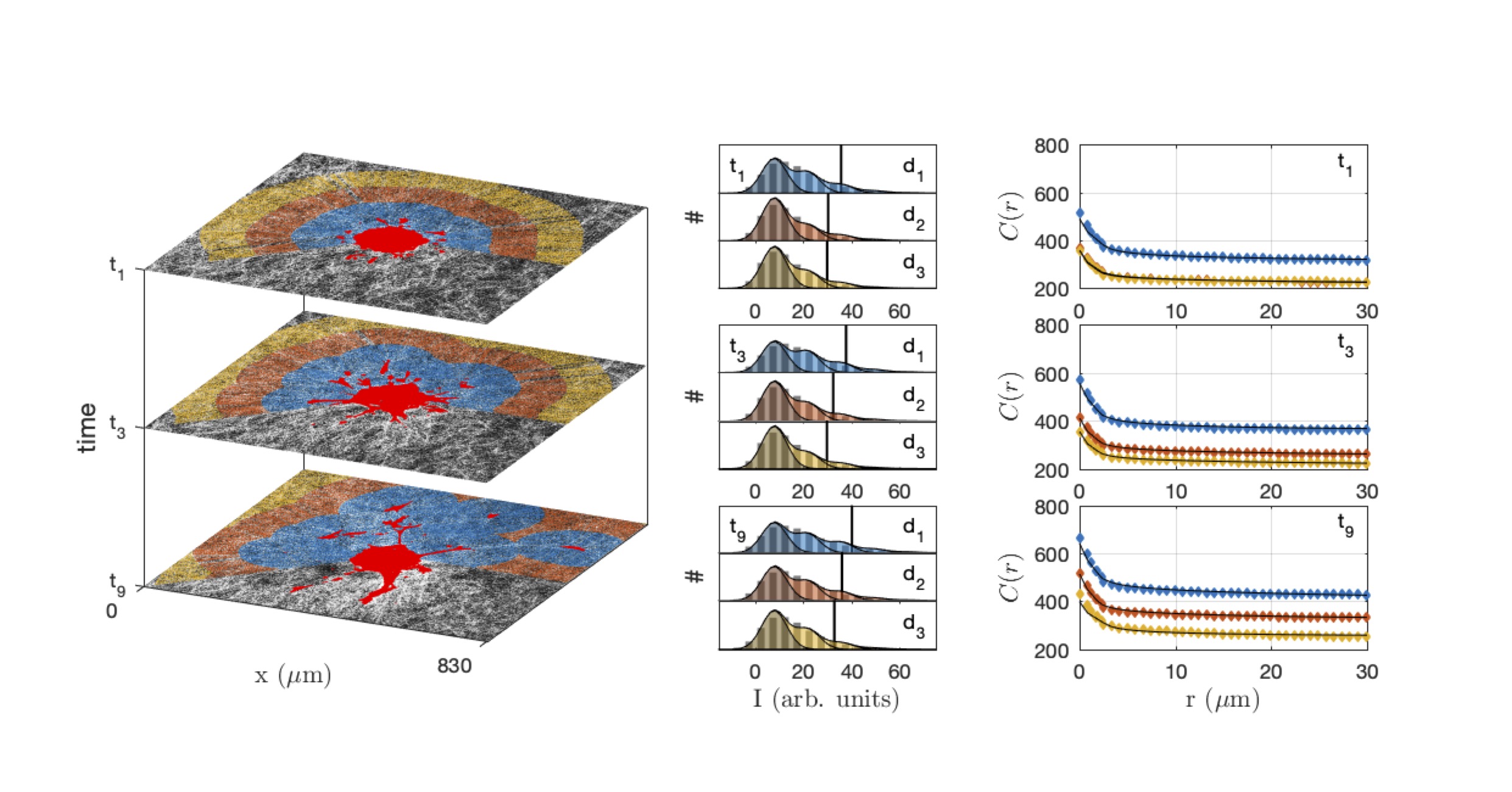}
\caption{Collagen structure analysis in the neighbourhood of the CAF spheroid, at different times ($t_1 = 30$ min, $t_3 = 210$ min $t_9 = 750$ min) and distances from the cells ($d_1 < 100 \ \mu$m, $100 \ \mu\textrm{m} < d_2 < 200 \ \mu\textrm{m} $ and $200 \ \mu\textrm{m} < d_3 < 300 \ \mu\textrm{m} $). In the original images (left) the gel is shown in grey and the cells in bright red, and the same colour code is used for the distances in the grey-tone distributions (middle) and correlation functions (right). The middle panel displays the distributions of grey-tone intensity $I$, with the bars being the experimental distributions and the 90 \% percentile shown as a vertical line. The shaded areas are the grey-tone model, with the background contribution (darker) and increasing number of fibre overlaps. In the right panel the grey-tone correlation functions $C(r)$ are shown, with the dots being the experimental values and the solid lines being the grey-tone model. }
\label{fig:spheroid_sampling}
\end{figure}

In order to quantitatively investigate the modification of the gel structure in relation with cell migration, the gel area surrounding the spheroid was decomposed into layers corresponding to increasing distances to the closest cells (shown as different colors in Fig. \ref{fig:spheroid_sampling} left). At each time step, the layers were recalculated according to the updated position of the cells. The shapes of the equal-distance layers become increasingly complex with time, as a consequence of the irregularity of the cell invasion pattern. The histogram of grey tones and the correlation function of the collagen were measured within each layer at each time step (Fig. \ref{fig:spheroid_sampling} middle and right). 

The densification of the fibres in the vicinity of the spheroid as well as of the migrating cells is manifest in Fig. \ref{fig:spheroid_sampling} through the progressive shifting of the grey-tone distributions towards brighter values (middle panel), and notably the 90 \% percentile (vertical lines). A similar evolution is observed in the correlation functions that progressively shift towards larger values with time. The evolution is quite rapid for the collagen in direct contact with the cells (d$_1$) and much slower for larger distance (d$_3$). 

\begin{figure}
\centering
\includegraphics[width=14cm]{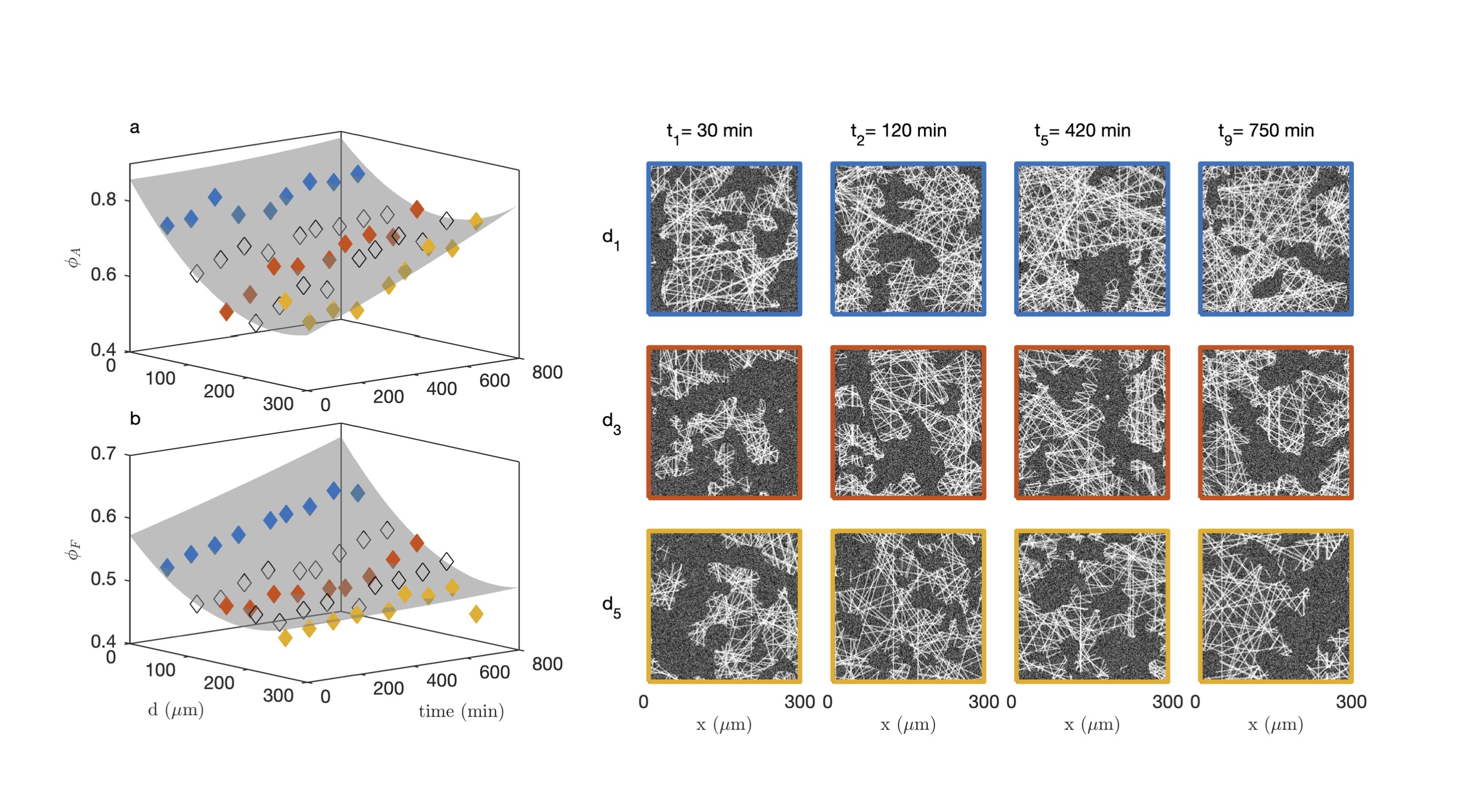}
\caption{Fitted collagen structure for various times and distances to the CAF cells, with aggregate density $\phi_A$ (a) and fibre density within the aggregates $\phi_F$ (b). The grey surface is a guide to the eye. Specific realizations are shown on the right panel to illustrate the significance of the parameters. The colour code for distances is the same as in Fig. \ref{fig:spheroid_sampling}; the non-colored symbols point to intermediate distances not shown in Fig. \ref{fig:spheroid_sampling}.}
\label{fig:spheroid_fit}
\end{figure}

In order to analyse the structural modifications corresponding to the changes in grey-tone distributions and correlation functions, all the data measured at 9 successive time steps and 5 different distances from the cells were fitted simultaneously with the grey-tone model (see Fig. \ref{fig:spheroid_sampling} for 3 times and 3 distances). Among the parameters of the grey-tone model, the densities and sizes bear a structural meaning (see Tab. \ref{tab:fitted_parameters}), but other parameters merely characterize the imaging mode. This is notably the case for the fibre contrast $\Delta$, the background intensity $b$, and the noise intensity $\sigma$ (see Eqs. \ref{eq:greytone_correlation} and \ref{eq:greytone_histogram}). For the fitting of the data, different values of the structural parameters were allowed for each time and distance, but a unique value of the imaging parameters was allowed for all images. This is justified as the time series images of the gel and spheroid were measured through a time-laps protocol with unchanged imaging parameters. The values of the imaging parameters from the fit are $b \simeq 9$, $\Delta \simeq 15$ and $\sigma \simeq 5$. The structural parameters are $D_F \simeq 2$ $\mu$m and $L_A \simeq 19$ $\mu$m for all distances and times, and the volume fractions $\phi_A$ and $\phi_F$ are as shown in Fig. \ref{fig:spheroid_fit}a and \ref{fig:spheroid_fit}b.

Fig. \ref{fig:spheroid_fit} testifies to qualitatively different structural changes in the gel in direct contact with the CAF cells and distant from them. Close to the cells (blue series), a large number of aggregates have already formed before the first measurement time ($t < 30$ min). It remains relatively constant afterward with density $\phi_A \simeq 80$ \%, which can also be described in terms of a matrix porosity that remains constant around 20 \%. In the meantime, the fibre density within the aggregates increases steadily all throughout the experiment. This contrasts with the situation far from the cells (yellow series), where the aggregate density $\phi_A$ increases far more progressively and the fibre density within the aggregates $\phi_F$ remains altogether constant. These qualitatively different structural changes close and far from the cells are illustrated with right panel of Fig. \ref{fig:spheroid_fit} for various distances and times.

\section{Discussion}

The structure of collagen gels is a highly hierarchical one \cite{Sun:2021}. Starting form molecular dimensions, the smallest structures are the tropocollagen triple helices (diameter 1.5 nm), assembled into protofibrils and fibrils (6 to 100 nm), woven into fibres (1 $\mu$m), which are eventually connected to form the gel matrix. The matrix, however, is not homogeneous at a scale larger than 10 $\mu$m, as visible in all images of Fig. \ref{fig:gel_images}. A central structure of collagen at large scale is he presence of fibre aggregates, which can also be thought of in terms of pores in the matrix. The latter are the most salient structures in the gels at the scale of the cells (Fig. \ref{fig:spheroid_images}), and it is therefore natural to enquire how they are modified by neighboring cells. We addressed this question by developing first a novel image analysis methodology to characterize the large-scale structure of collagen, by validating it on two acellular gels, and finally applying it to characterise the collagen structure surrounding a spheroid of CAF cells.

The characterisation of the disordered pore-and-aggregate structure of the gels is challenging, and calls for a statistical description. The approach we explored here is based on a mathematical modelling, which aims at statistically capturing the main structural characteristics of the gels through a small number of meaningful parameters. It is important to stress that the goal of the models is not to reproduce a local gel structure, as one would typically observe in a single image, but also to capture the structure variability. This feature is essential to assign a single value to the structural parameters based on many different images of one and a single gel. 

In that spirit we first considered a homogeneous fibre model (Fig. \ref{fig:fibre_model}). This simple model was found to be incompatible the experimental covariances of the collagen, which statistically confirmed the presence of fibre aggregates and pores. We then proposed a two-scale model, where the small-scale structure is described as a homogeneous model of random fibres, and the large-scale aggregate structure is obtained by using a clipped Gaussian-field to carve pores out of it Fig. \ref{fig:model_aggregates}). In this two-scale model, two parameters $\phi_A$ and $L_A$ describe the density and size of the aggregates, and two parameters $\phi_F$ and $D_F$ describe the density and the diameter of the fibres inside the aggregates. Analytical expressions were derived to fit experimental characteristics of the gels measured from the images, and infer the values of the mentioned structural parameters. 

A classical image analysis method consists in segmenting first the images to identify objects of interest and subsequently measure them. This approach was proved unsuitable in the present context, as segmentation would result in the loss of valuable structural information. The collagen-rich areas of the gels are indeed characterized by broad grey-tone distributions (see Fig. S1b), with large intensities being associated with the close proximity or overlapping of many fibres. Image segmentation would amount to treating all these regions on the same footing, and would therefore bias any measurement. 

We developed a method to identify the model parameters directly from grey-tone microscopy images, through the fitting of the covariance function and histogram of intensities. The covariance -or correlation function, in the case of grey-tone images- is a very informative structural descriptor \cite{Jiao:2011,Gommes:2012} that can easily be measured on any image through fast Fourier transforms. Covariance analysis offers a variety of advantages. First, it provides a robust and fully objective statistical description of the spatial distribution of the objects that make up an image. This is particularly useful for structures as complex and disordered as fibrillar collagen gels. Second, it reduces the impact and possible bias of image preprocessing, which is absent altogether in the case of grey-tone correlation functions. Finally, it also provides an efficient way to cope with noisy images, as noise is typically defined as the non-correlated contribution to an image. 

Globally, the procedure based on grey-tone measurements (correlation functions and intensity distributions) and on their modelling proved to be quite consistent with respect to the two types of imaging modes considered in the paper. The absolute values of the parameters obtained through reflectance and fluorescence data differ slightly, but identical trends are detected when comparing the 2mg/mL and 3mg/mL acellular gels (Tab. \ref{tab:fitted_parameters}). The overall fibre density $\phi_1$ is larger in the 3 mg/mL gel, as it should. Our analysis shows that this is due largely to an increased density of aggregates $\phi_A$ while the fibre density within the aggregates $\phi_F$ is almost identical in the two gels. The characteristic size of the aggregates $L_A$ is larger in the 3mg/mL gel because the aggregates are more numerous and form a larger connected structure. To further highlight the importance of using grey-tone images, we also performed the same analysis on segmented images. In that case, segmentation leads to the undesirable situation where the density of the segmented images is opposite to the known collagen concentration of the gels in the case of fluorescence data (Tab. \ref{tab:fitted_parameters} and Fig. \ref{fig:gel_covariance}a). 

From a methodological point of view, the application of the grey-tone model to analyze the structure of the 2mg/mL and 3 mg/mL gels in both reflectance and fluorescence modes, validated the overall image analysis approach and enabled us to apply it to the dynamic model of CAF spheroid invasion (Figs. \ref{fig:spheroid_sampling} and \ref{fig:spheroid_fit}). Matrix remodelling has been widely documented in earlier works \cite{Balcioglu:2016,Kopanska:2016,Piotrowski:2017,Winkler:2020,Mark:2020}. Contractility of collagen-embedded cells generates strains on the collagen fibres, leading to their local densification  \cite{Doyle:2015,Doyle:2021} and reorganization \cite{Kopanska:2016,Piotrowski:2017,Chen:2019}. This process is especially evident for cells of mesenchymal origin such as CAFs, which are characterized by high expression levels of the contractile myosin protein and collagen-binding integrins \cite{Doyle:2021}. In our experimental model, CAF spheroids were treated with PDGF-BB, a growth factor known to stimulate collagen gel contraction \cite{Gullberg:1990,Reyhani:2017}. By pulling on collagen fibers, cells stiffen their microenvironment and induce irreversible collagen deformations, thereby altering locally the mechanical properties of the ECM. This remodeled microenvironment in turn provides guidance cues for the nearby cells through durotactic and/or topotactic migrations \cite{Sengupta:2021,Lo:2000,Duchez:2019}. Besides these guidance cues, high concentrations of fibrillar collagen also promote the local invasion of cells by inducing the formation of specialized cellular extensions termed invadopodia implicated in the local proteolytic remodeling of the ECM \cite{Artym:2015}.

Our use of a two-scale stochastic model to describe the collagen structure and the identification of its parameters from grey-tone images, enables us to describe the ECM remodelling in more detail than simply the collagen density. The pristine state of the ECM in the CAF spheroid model is the one observed at early times $t$ and large distances $d$ from the cells. The relevant values of the aggregate and fibre densities are $\phi_A \simeq 0.6$ and $\phi_F \simeq 0.5$ based on Figs  \ref{fig:spheroid_fit}a and  \ref{fig:spheroid_fit}b. These values are reasonably consistent with those of the 2mg/mL acellular gels in Tab. \ref{tab:fitted_parameters}, and the slight differences can be attributed to the lower resolution used in the time-resolved analysis of the spheroid. Starting from that initial state, our space- and time-dependent analysis of the collagen structure shows that the ECM densification happens via two distinct mechanisms, namely: the increased density of fibre aggregates $\phi_A$ and the increased fibre density $\phi_F$ within the aggregates. The two mechanisms are both at work in the gel in close contact with the cells. The aggregate density increases very fast, from about $\phi_A \simeq 60 $\% to 80\% in less than 30 mins (Fig. \ref{fig:spheroid_fit}a), similar to the cancer cell spheroids investigated by \cite{Chen:2019} which exert a strong contraction of the surrounding collagen immediately after embedding in the matrix. The fibre densification also takes place inside the aggregates is a much more progressive way, as it occurs over hours (Fig. \ref{fig:spheroid_fit}b). 

The global densification of the gel resulting from the two processes is quite significant, as the overall fibre density in contact with the cells - estimated as $\phi_1 = \phi_A \times \phi_F$ - approximately passes from $\phi_1 \simeq $ 30\% to 50\%. It is important to stress that this only concerns the fraction of fibres that are visible in our experiments, which is determined by the imaging technique (CRM vs CFM) and by the spatial resolution of the imaging system ($\sim$0.59$\mu$m/pixel in our case). As smaller fibrils and protofibrils are undetected at the considered scale, the apparent densification does not contradict the fact that the total collagen concentration has to remain locally constant. We also cannot exclude that a fraction of the aggregate densification observed by CRM in our spheroid model results from a CAF-mediated partial reorientation of the more vertical collagen fibres relative to the imaging plane \cite{Jawerth:2010}. In any event, the analysis testifies to significant yet different remodelling of the gel structure at two scales simultaneously.

Interestingly, the remodelling of the collagen is not limited to the regions in direct contact with the cells. Significant structural changes are observed also in regions as far as 300 $\mu$m away from the closest CAF cell, but the phenomenology is distinctly different there. In those regions, the density of fibre aggregates increases slowly, over hours, while the fibre density within the aggregates remains constant. Similar long-range (up to 1300 $\mu$m) mechanical signals generated by fibroblasts have been shown to propagate through fibrillar collagen networks. Such mechanical cues can be sensed far beyond the signal source by cells sharing the same substrate, which is responsible for the ability of contractile fibroblasts to induce the migration of macrophages towards the force source \cite{Pakshir:2019}.

From a methodological point of view, it is important to stress that the discrimination between the two types of densification processes is robust. Considering Fig. \ref{fig:model_grey}a$_2$, an increase of aggregate density $\phi_A$ with constant fibre density $\phi_F$ would be manifest through a reduction of the background contribution (dark blue) in favor of the fibre contribution (bright blue) but the shapes of the two contributions would remain unchanged. By contrast, any increase in the fibre density $\phi_F$ is accompanied by an increasing number of pixels where the fibres overlap, which would necessarily extend the grey-tone distribution towards higher intensities. With that in mind, the observed shifting of the 90\%-percentile of the image intensity in Fig. \ref{fig:spheroid_sampling} can only be interpreted as an increase of $\phi_F$, independently of any evolution of the aggregate density $\phi_A$. As for the grey-tone correlation function, its two asymptotic values for $r=0$ and $r \to \infty$ depend only on the total fibre density $\phi_1 = \phi_A \times \phi_F$. The combination of the grey-tone distribution and correlation function is therefore key to discriminate the two different densification processes.

\section{Conclusion}

In this study, we developed an innovative image analysis approach to investigate the remodelling of fibrillar collagen in a 3D spheroid model of cellular invasion. Unlike existing work, most of which focus on fibre the densification of the collagen network and on small-scale fibre reorientation, we focused here on the structural modification of the collagen matrix at the scale of a few microns, comparable with that of the cells. This was achieved by first developing a novel image analysis method based on the stochastic modelling of the acellular gel structure, and applying it afterwards to study the space- and time-dependent reshaping of the collagen matrix by migrating CAFs.

The analysis of acellular collagen gels (without embedded cells) investigated by confocal microscopy in both reflectance and fluorescence modes, shows that the structure of the gels is not homogeneous at the scale of about 10 microns. The structure consists in regions with high fibre density separated by depleted regions, which can be thought of as fibre aggregates and pores. In order to mathematically describe this structure, we developed a two-scale stochastic model with a clipped Gaussian-field model for the aggregates and pores, and a homogeneous Boolean model to describe the fibre network within the aggregates. We also developed a method to identify the model parameters from the grey-tone distributions and correlation functions of the gel images. The specificity of the method is that it applies to the unprocessed grey-tone images, and it can therefore be used with noisy time-lapse reflectance images of non-fluorescent collagen.

When applied to the collagen-embedded CAF spheroid images, the developed method testifies that the invasion of PDGF-BB-treated CAFs is accompanied by an overall densification of the collagen gel while the sizes of the pores and aggregates, as well as that of the fibres, remains largely unchanged. Interestingly, the densification occurs differently for the gel in direct contact with the cells or far away from them. The gel in close contact with the invading cells densifies through the rapid increase of the number of aggregates, over less than 30 min, followed by the slow increase of fibre density within the aggregates, over hours. By contrast, the densification occurring in the gel located farther away from the cells occurs via the slow increase of the aggregate density, while the density of fibres within the aggregates remains constant.

At the present stage, one can only speculate on the biomechanical mechanisms responsible for the two-scale densification. The very observation of two distinct phenomenologies hint at diverse mechanisms, which presumably involve both biochemical and mechanical effects. 

%
%

\section*{Conflict of Interest Statement}

The authors declare that the research was conducted in the absence of any commercial or financial relationships that could be construed as a potential conflict of interest.

\section*{Author Contributions}

AN and EM designed the biological experiments; CJG and SB designed the mathematical analysis; IB prepared the biological samples; TL performed the confocal microscopy imaging; CJG developed the mathematical models, analyzed the data and prepared all figures; CJG, EM, SB and AN wrote the manuscript.

\section*{Funding}

CJG and EM are grateful to the Funds for Scientific Research (F.R.S.-FNRS, Belgium) for Research Associate positions.  This work was supported by FNRS-T\'el\'evie grants 7.4589.16 and 7.6527.18.

\section*{Acknowledgments}
Technical support from the GIGA Imaging platform of the Universit\'e de Li\`ege is gratefully acknowledged.

\section*{Supplemental Data}

A Supplementary Material file is available with details about the method used to segment the gel images into fibre and background pixels, and about the Gaussian Random Field approach used to model the fibre aggregates.


\bibliography{gelmodel}

\end{document}